\documentclass[12pt]{iopart}
\usepackage{afterpage}
\usepackage{color}
\usepackage{booktabs}

\usepackage[colorlinks=true, linkcolor=blue, citecolor=blue, urlcolor=blue]{hyperref}
\usepackage{subfig} 
\usepackage{graphicx} 
\usepackage[format=plain,justification=justified]{caption} 
\usepackage{orcidlink}
\begin{document}
\title[Control of Band Alignment and Mechanical Anisotropy in  MoS$_2$-Organic Hybrids]{Chemical Control of Mechanical Anisotropy  and Band Alignment in Perylene-based Two-dimensional MoS$_2$-Organic Hybrids}

\author{Mohammed El Amine Miloudi \orcidlink{0000-0002-2654-8758},
Oliver Kühn \orcidlink{0000-0002-5132-2961}}
\address{University of Rostock, Institute of Physics, Albert-Einstein-Str. 23-24, D-18059 Rostock, Germany }
 \ead{oliver.kuehn@uni-rostock.de}
\vspace{10pt}
\begin{indented}
\item[]\today
\end{indented}

\begin{abstract}
This study presents a comprehensive investigation of hybrid interfaces formed by monolayer MoS$_2$ coupled with the organic molecules perylene (P), perylene diimide (PDI), and perylene orange (PO). Using density functional theory, we demonstrate the extent to which the mechanical and electronic properties of a hybrid system can be altered by the chemical modification of a given chromophore. The three systems exhibit distinct differences due to their chemical composition and van der Waals contact enabled by their geometry. All systems are structurally stable. The binding energies follow the order PD$>$P$>$PO due to the large $\pi$-system (PD) and strong structural distortion (PO). Young's modulus and Poisson's ratio exhibit pronounced anisotropy in all cases. PO exhibits the greatest anisotropy due to steric effects and a permanent dipole, which introduce directionality to the molecule-surface interaction. Physisorption is accompanied by net charge transfer in the same order as the binding energies. The associated interfacial polarization results in a change in the work function compared to pristine MoS$_2$  in the order P$>$PO$>$PD. Finally, the presence of organic molecules introduces states into the MoS$_2$ energy gap, with the band alignment being either type II (P, PO) or type I (PD).
\end{abstract}

\maketitle

\section{Introduction}

Transition metal dichalcogenides (TMDs) have emerged as key materials in  research due to their tunable layered structures and exceptional electronic properties. With the general formula MX$_2$ (where M is a transition metal and X is a chalcogen atom), TMDs exhibit a wide range of electronic behaviors, from insulating to metallic states. Their "sandwich-like" structure, a result of strong in-plane covalent bonding and weaker van der Waals (vdW) forces between layers, combined with the ability to modify layer thickness easily, makes TMDs promising candidates for innovations in optoelectronics, photovoltaics, and even space technology \cite{liang69_1031,wilson69_193,lee76_,murray72_746,kasowski73_1175}.

Among the TMD family, semiconducting variants such as MoX$_2$ and WX$_2$ (where X = S, Se, Te) stand out due to their tunable band gaps, which distinguish them from other 2D materials like graphene. This tunability makes TMDs highly adaptable for nanoelectronics applications \cite{coehoorn87_6195,kang17_1610}. However, a key challenge for integrating TMDs into optoelectronic devices is the dependence of their electronic properties on layer thickness \cite{kang16_597,villaos19_1,husain22_064422,lin16_165203,zhang16_29615,kim21_085404}.

Molybdenum disulfide (MoS$_2$), a prototypical semiconducting TMD, is particularly notable due to its scalable production and remarkable optoelectronic properties \cite{schmidt14_1909,perea-lopez14_011004,chen18_1690,ermolaev20_1,brill21_32590}. Its band structure gives rise to distinct exciton species, including A and B excitons resulting from spin-orbit splitting of the valence band, and a non-emissive C exciton. These excitons, which have large binding energies and exhibit significant delocalization, show observable shifts in their photoluminescence (PL) spectra as the number of layers varies \cite{Splendiani.2010}. Properties like quantum confinement, valley polarization, tunable ferroelectricity, and robust light-matter interactions further enhance MoS$_2$’s appeal for optoelectronic applications \cite{steinhoff15_6841,trolle14_235410,schwandtkrause2025ferroelectric-d55,wang20_1749}.

Parallel to the development of TMDs, the field of organic semiconductors has  advanced rapidly. Organic materials are renowned for their strong light absorption, cost-effectiveness, and compatibility with flexible substrates \cite{huang18_3241, zarrabi23_3174, li10_529, gupta15_8468}. A particularly promising area is the integration of TMDs with organic semiconductors, leading to organic/inorganic hybrid systems with potentially improved optoelectronic properties \cite{Draxl.2014}. Hybrid systems combining MoS$_2$ with organic molecules, such as 9-(2-naphthyl)-10-[4-(1-naphthyl)phenyl]anthracene (ANNP) \cite{declercq23_11260}, vanadyl phthalocyanine (VOPc) \cite{kong22_1253}, and tin (IV) phthalocyanine dichloride (SnCl$_2$Pc) \cite{kong22_1253} exhibit improved interfacial charge transfer (CT) and PL performance. For instance, the MoS$_2$/VOPc heterostructure shows significant quenching of MoS$_2$ PL due to efficient CT from VOPc to MoS$_2$, generating interlayer excitons via mid-gap states.

\begin{figure}[t]
  \centering
  \includegraphics[width=0.8\textwidth]{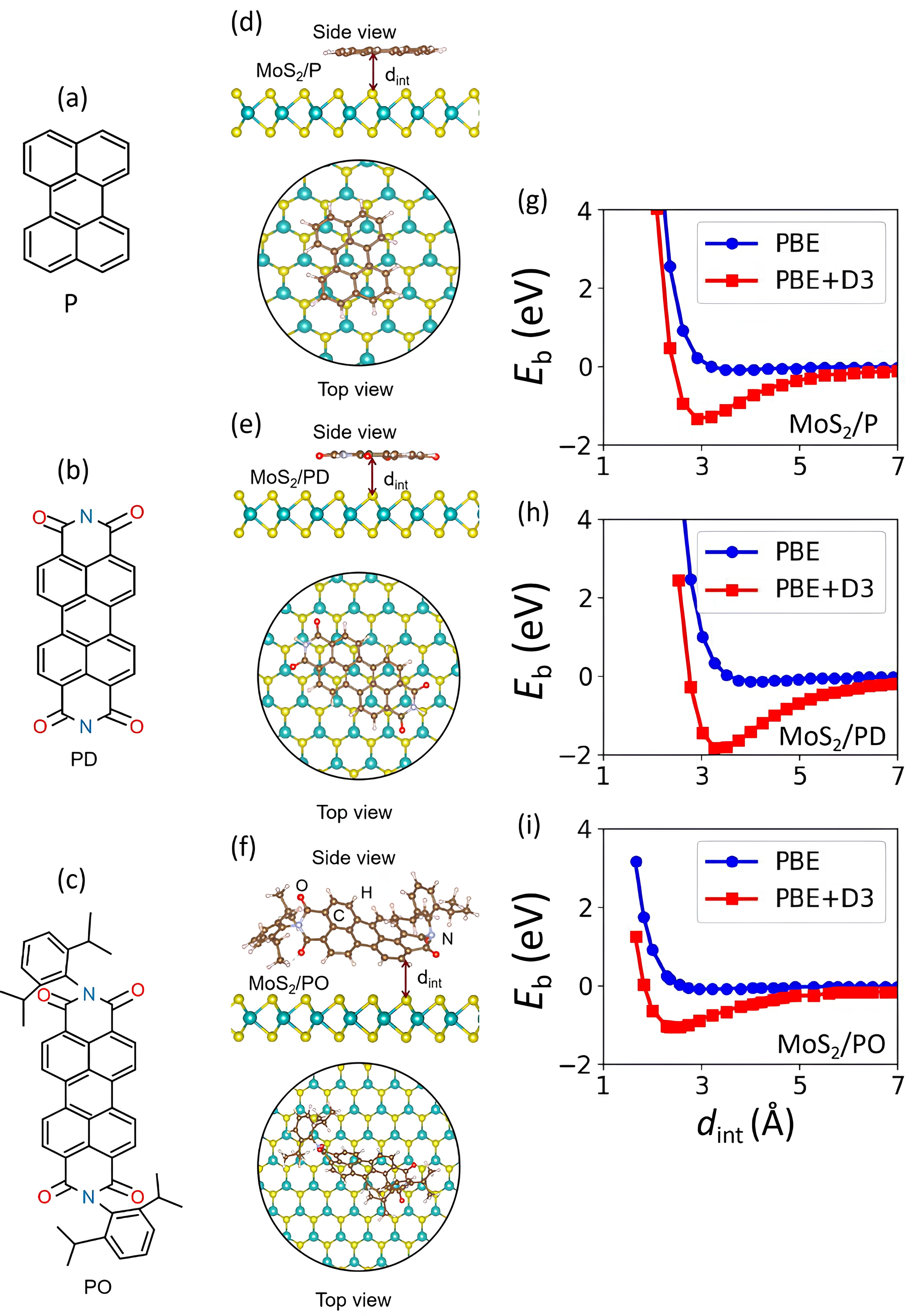}
  \caption{Chemical formulas of (a) P (C$_{20}$H$_{12}$), (b) PD (C$_{24}$H$_{10}$N$_2$O$_4$), and (c) PO (C$_{48}$H$_{42}$N$_2$O$_4$). Side and top views of optimized (d) MoS$_2$/P, (e) MoS$_2$/PD, and (f) MoS$_2$/PO hybrid interfaces.  Binding energy profiles of (g) MoS$_2$/P, (h) MoS$_2$/PD, and (i) MoS$_2$/PO as functions of interlayer distance, obtained using the PBE and PBE+D3 models.}
  \label{1}
\end{figure}

In organic electronics and high-performance pigment technology, the perylene family plays a pivotal role due to its exceptional optical and electronic properties \cite{harding20_2959,powell22_2026,moulin14_,huang11_2386,Wurthner.2016}. Perylene derivatives can be customized with various substituents to fine-tune their characteristics. Prominent examples are perylene diimide  and perylene orange as shown in Fig.~\ref{1}. Known for their vivid colors, high photostability, and excellent electronic performance, perylene derivatives are widely used in organic solar cells, organic light-emitting diodes, bioimaging, and sensing applications \cite{guthmuller09_154302,gavrila04_4657,chand21_6486,kaur13_502,obaidulla20_1901197,zong18_9532,kumar23_126964}. 
    
The interaction between MoS$_2$ and perylene-based organic semiconductors results in particularly rich optoelectronic behaviors.   For instance, the MoS$_2$/PTCDA heterostructure demonstrates a substantial increase in PL intensity, driven by strong interfacial interactions and crystalline ordering of the PTCDA layer. This behavior leads to a significant PL peak shift, reflecting robust coupling at the interface that can be exploited to tune optoelectronic properties \cite{kong22_1253,habib18_16107,Habib.2020}.
In a recent study, the PL of MoS$_2$/perylene orange interface was investigated \cite{volzer23_3348a}. It was found that molecular PL is quenched by efficient interfacial charge separation. This was in accord with band structure calculations, pointing to a type II band alignment \cite{Miloudi.2024}. In Ref. \cite{Miloudi.2024} it was further shown that the type of alignment depends on the applied strain, i.e. upon compression of the MoS$_2$/perylene orange interface a transition to type I alignment was proposed. Controlled application of strain thus may provide a means for tuning interfacial properties of hybrid systems.

Considering organic/inorganic interfaces it is frequently highlighted that chemical design of the organic part provides a high degree of flexibility for tuning mechanical and optoelectronic properties. Here we explore, using Density Functional Theory (DFT), to what extent key interfacial properties can be tuned for a given class of chromophores. As a specific example we will use perylene (P) and its derivatives perylene diimid (PD) and perylene orange (PO). 
Specifically, we  focus on mechanical and electronic properties of  the hybrid interface. By examining band alignment, work function variation, and charge density distribution, we specifically aim to elucidate the molecular-level interactions that govern the optoelectronic behavior of these hybrid systems.  

\section{Computational Details}
\label{sec:comp}

First-principles calculations were performed using DFT with the projector augmented-wave (PAW) method and the Perdew-Burke-Ernzerhof (PBE) generalized gradient approximation (GGA) functional, as implemented in the Vienna ab initio Simulation Package (VASP) \cite{kresse96_15,kresse96_11169,blochl94_17953,perdew96_3865}.  To prevent interlayer interactions, a vacuum layer of 20 \AA{} was added along the $z$-axis, with dipole corrections applied to mitigate spurious interactions between periodic images. A $9 \times 9 \times 1$ supercell of MoS$_{2}$ was employed to model interactions between the MoS$_{2}$ monolayer and the organic molecules. Note that we consider the limit of low coverage, i.e. only a single organic molecule per supercell is taken into account. We performed geometry optimization starting from parallel and perpendicular orientations with the molecules being initially in gas phase geometry. Energy and force convergence thresholds were set to $10^{-6}$ eV and 0.01 eV/\AA, respectively. Computational parameters, including plane-wave cutoff energy (450 eV), smearing width (0.05 eV), and $k$-point density ($1 \times 1 \times 1$), were optimized to achieve a balance between precision and computational efficiency.

DFT exchange-correlation functionals, such as those based on the Local Density Approximation (LDA) \cite{chen18_1690} and the GGA \cite{harding20_2959}, are generally effective in describing covalent and ionic bonding. However, these functionals are not appropriate for systems involving weak vdW interactions, as they do not explicitly account for such forces. For example, GGA functionals like PW91 and PBE \cite{harding20_2959} fail to accurately describe interactions between layered materials, such as h-BN or graphene, or between these layers and transition-metal (111) surfaces \cite{guthmuller09_154302,blochl94_17953}. In systems involving MoS$_2$ and organic molecules, the relevance of vdW interactions is not immediately clear. To account for these interactions, the empirical PBE+D3 method was employed, as opposed to the computationally more demanding vdW-DFT approach \cite{gavrila04_4657, chand21_6486}.

To determine the binding energies ($E_{\rm b}$) of MoS$_{2}$/organic hybrid interfaces, the following equation was used

\begin{equation}
    E_{\rm b} = E_{\rm{MoS}_2/\rm{organic}} - E_{\rm{MoS}_2} - E_{\rm{organic}}\,.
\end{equation}

Here, \(E_{\rm{MoS}_2/\rm{organic}}\) is the total energy of the MoS$_{2}$-organic composite system, \(E_{\rm{MoS}_2}\) represents the energy of the isolated MoS$_{2}$ monolayer, and \(E_{\rm{organic}}\) refers to the energy of the isolated organic molecule. The geometries of all subsystems are geometry-optimized.

Additional electronic structure calculations for isolated molecules were performed at the DFT/PBE level with a 6-311G(d,p) basis set using the Q-Chem 5.3 package \cite{Shao.2015}.

To systematically investigate mechanical property modulations in a monolayer of MoS$_2$ and its composite interfaces with perylene-based molecules, in-plane uniaxial strains were applied. The strain magnitude (\(\varepsilon\)) was characterized by the alteration in the lattice parameter, defined as

\begin{equation}
    \varepsilon = \frac{100\% \times (a - a_0)}{a_0}\, ,
\end{equation}

where \(a_0\) and \(a\) represent the lattice constants of the unstrained and strained systems, respectively.
The focus centered on the analysis of planar elastic stiffness coefficients, specifically \(C_{11}\), \(C_{12}\), and \(C_{22}\). These coefficients were extracted by fitting the  supercell's energy, \(U\), for certain values (\(\epsilon_{11}\), \(\epsilon_{22}\)).

The elastic stiffness coefficients were computed as follows

\begin{eqnarray}
    C_{11} &=& \frac{1}{A_{0}}\frac{\partial^2 U}{\partial \epsilon_{11}^2}, \\
    C_{12} &=& \frac{1}{A_{0}}\frac{\partial^2 U}{\partial \epsilon_{11} \partial \epsilon_{22}},
\end{eqnarray}

where \(A_0\) is the equilibrium (zero-strain) lateral area of the supercell used to model the MoS$_2$/organic interface.
For a hexagonal lattice, \(C_{11}\) is inherently equal to \(C_{22}\). The Young's modulus (\(Y\)), shear modulus (\(G\)), Poisson's ratio (\(\nu\)), and bulk modulus (\(K\)) were calculated using the following relationships

\begin{equation}
Y = \frac{C_{11}^2 - C_{12}^2}{C_{11}}, \quad G = \frac{C_{11} - C_{12}}{2}, \quad 
\nu = \frac{C_{12}}{C_{11}}, \quad K = \frac{C_{11} + C_{12}}{2}.
\end{equation}

In addition, the angular-dependent material properties were evaluated using the following equations, valid for a hexagonal lattice, to get Young's modulus ($Y(\theta)$) and Poisson's ratio ($\nu(\theta)$)

\begin{equation} Y(\theta) = \frac{C_{11}^2 - C_{12}^2}{C_{11} + C_{12} + (C_{11} - C_{12}) \cdot \cos(2 \theta)}, 
\label{eq:Y}
\end{equation}

\begin{equation} \nu(\theta) = \frac{C_{12} + (C_{11} - C_{12}) \cdot \cos(2 \theta)}{C_{11} + C_{12} + (C_{11} - C_{12}) \cdot \cos(2 \theta)}. 
\label{eq:nu}
\end{equation}

The strain range spans from \(-8\%\) to \(8\%\), with a step size of 0.02, elucidating detailed changes in the mechanical and structural properties of the MoS$_2$ monolayer and its hybrid configurations under controlled strain conditions. To evaluate whether the strain remains within the elastic limit, the per-atom strain energy ($E_{\rm S}$) was calculated using

\begin{equation}
    E_{\rm S} = \frac{1}{n} \left( U_{\rm{strained}} - U_{\rm{unstrained}} \right),
    \label{eq:es}
\end{equation}

where \( n \) is the total number of atoms in the simulation supercell. This normalization allows comparison of strain energy on a per-atom basis, independent of system size.

Obtaining accurate band alignment for organic/inorganic hybrid structures using DFT is, in general, a challenging task. Polarization effects are not well described in Kohn-Sham theory and more accurate fully self-consistent many-body perturbation theory (GW) is computationally not feasible and one has to resort to the G$_0$W$_0$ approximation \cite{Oliva2022}. System-specific nonempirically tuned range-separated hybrid functionals, that show very good performance for isolated molecules \cite{Bokarev.2015zii}, are in general difficult to adjust \cite{krumland_2300089}. Krumland et al. suggested a pragmatic method  based on the subsystem idea \cite{krumland21_224114,krumland_2300089}. Applying this model to phthalocyanine on MoS$_2$ it has been found that a good approximation for the band alignment is obtained from PBE and PBE0  DFT calculations of the isolated subsystems \cite{krumland24_5350}. In case of  the MoS$_2$/PO interface in Ref. \cite{Miloudi.2024} we have found decent agreement with the PBE result. Hence we will assume that PBE is reliable also for the two other perylenes investigated in the present study.

At the MoS$_2$/organic interface, charge transfer modifies the electrostatic potential profile, leading to shifts in vacuum levels and work functions. This built-in potential drives electron flow until Fermi level alignment is achieved, establishing equilibrium. The resulting interfacial electric field influences carrier dynamics, band bending, and exciton dissociation.
To assess CT effects and the resulting electrostatic potential alignment at the MoS$_2$/organic interfaces, we computed the planar-averaged charge density difference  for the electronic ground state. These quantities allow visualization and quantification of electronic charge redistribution due to interfacial interactions and their electrostatic consequences.

The three-dimensional charge density difference is defined as
\begin{equation} \label{eq:deltarho}
    \Delta \rho(\mathbf{r}) = \rho_{\mathrm{MoS}_2/\mathrm{Organic}}(\mathbf{r}) - \rho_{\mathrm{MoS}_2}(\mathbf{r}) - \rho_{\mathrm{Organic}}(\mathbf{r}),
\end{equation}
where $\rho_{\mathrm{MoS}_2/\mathrm{Organic}}(\mathbf{r})$ is the total charge density of the hybrid system, and $\rho_{\mathrm{MoS}_2}(\mathbf{r})$ and $\rho_{\mathrm{Organic}}(\mathbf{r})$ correspond to the isolated MoS$_2$ monolayer and organic molecule, respectively, computed in the same supercell geometry without electronic interaction.

To analyze the charge redistribution perpendicular to the interface (i.e. in stacking direction), we perform a planar average over the $xy$ plane:
\begin{equation} \label{eq:deltarhoz}
    \Delta \rho(z) = \frac{1}{A_0} \int\int \Delta \rho(x,y,z)\, dx\, dy \, .
\end{equation}
 
Finally, we also report on the  work function, $\Phi$, of a surface given by
\begin{equation}
    \Phi = E_{\rm v} - E_{\rm F},
\end{equation}
where $E_{\rm v}$ is the electrostatic potential in the vacuum region (i.e., the vacuum level), and $E_{\rm F}$ is the Fermi energy relative to the internal potential of the material.

\section{Results and Discussion}
The three considered molecules have different structural and thus electronic characteristics that are pertinent to the interfacial binding.
Comparing the planar P and PD we notice not only the heteroatoms in PD, but also its larger $\pi$-electron system providing an increased vdW contact area with the MoS$_2$ surface.
PO also consists of seven fused rings, but its bulky side groups make it notably non-planar and provide steric hindrance for the interaction with the surface. In what follows, we will present and discuss results, which highlight how these differences are reflected in structural, mechanical and electronic properties of the hybrid interfaces. 
\subsection{Structure and Binding Energy}
The structural properties of the MoS$_2$/organic hybrid interfaces were investigated by considering both parallel and perpendicular molecular orientations. Figures~\ref{1}(a)–(c) show the chemical formulas of the organic molecules P, PD, and PO, while Figs.~\ref{1}(d)–(f) present the side and top views of the geometry-optimized MoS$_2$/P, MoS$_2$/PD, and MoS$_2$/PO hybrid interfaces. The perpendicular configurations are shown in the Supplemental Material (Suppl. Mat.), Fig.~S1. Due to the missing vdW contact between the $\pi$-system and the MoS$_2$ surface, their binding strength is much reduced compared to the parallel cases.

The calculated in-plane lattice constants for pristine MoS$_2$, MoS$_2$/P, MoS$_2$/PD, and MoS$_2$/PO are 3.18, 3.16, 3.16, and 3.17~\AA, respectively. This minimal variation suggests that adsorption induces only weak structural perturbation, indicative of physisorption governed by vdW interactions. The absence of a commensurate overlayer further supports the non-epitaxial nature of the interface. The slightly reduced lattice constants in MoS$_2$/P and MoS$_2$/PD may result from enhanced vdW interactions due to the planar geometry of these molecules, enabling closer contact with the MoS$_2$ surface. In contrast, the marginally larger lattice constant in MoS$_2$/PO is likely attributable to steric repulsion arising from the non-planar molecular structure.

Figure~\ref{1}(g-i) shows the binding energy as a function of interlayer distance (cf. panels (a-c) for the definition of $d_{\rm int}$). The equilibrium interlayer distances for MoS$_2$/P, MoS$_2$/PD, and MoS$_2$/PO are 2.91~\AA, 3.27~\AA, and 2.37~\AA, respectively, as summarized in Table~\ref{tab1}. Comparing results obtained with and without the D3 dispersion correction, it is evident that all systems gain significant stabilization from vdW interactions. Regarding the binding strength among the three molecules, the trend follows the available contact area between the $\pi$-system and the MoS$_2$ surface, i.e.  MoS$_2$/PD ($-1.83$~eV) $>$ MoS$_2$/P ($-1.32$~eV)$>$ MoS$_2$/PO ($-1.07$~eV). This trend reflects both the planar nature of PD and P molecules, which facilitates stronger  vdW interactions, and the more sterically hindered, non-planar geometry of PO, which limits close contact with the substrate. Regarding the lateral potential energy landscape, the physisorptive nature dominated by vdW forces suggests a relatively smooth and shallow corrugation. Consequently, the molecules are expected to experience, compared to the binding energy, low energy barriers for lateral diffusion on the MoS$_2$ surface, enabling mobility at finite temperatures. However, in particular for PO variations in molecular geometry and local adsorption sites may induce modulations of the lateral potential, warranting detailed exploration through explicit calculations of lateral energy barriers to fully characterize potential diffusion pathways (e.g. \cite{Coutre2025}).

\begin{table*}[ht]
\caption{Elastic constants (C$_{11}$, and C$_{12}$), Young modulus ($Y$), shear modulus ($G$), Poisson’s ratio ($\nu$), and bulk modulus ($K$) of MoS$_2$, MoS$_2$/P, MoS$_2$/PD, and MoS$_2$/PO.}
\begin{tabular}{l|cccccccccc}
  System   & $C_{11}$ (N/m) & $C_{12}$ (N/m)  & $Y$ (N/m) & $G$ (N/m) &  $\nu$ & $K$ (N/m)  \\
  \hline
MoS$_2$        & 138.5 & 31.37 & 131.4 & 53.6 & 0.23 & 85.2    \\
MoS$_2$/P    & 130.2 & 31.34 & 122.6 & 49.4 & 0.24 & 80.8 \\
MoS$_2$/PD    & 129.7 & 31.6 &  122.0 & 49.1 & 0.24 & 80.6 \\
MoS$_2$/PO    & 140.9 & 18.2 &  138.6 & 61.4 & 0.13 & 79.5   \\
\end{tabular}
\label{tab1}
\end{table*}

\begin{figure}[t]
  \centering
  \includegraphics[width=0.9\textwidth]{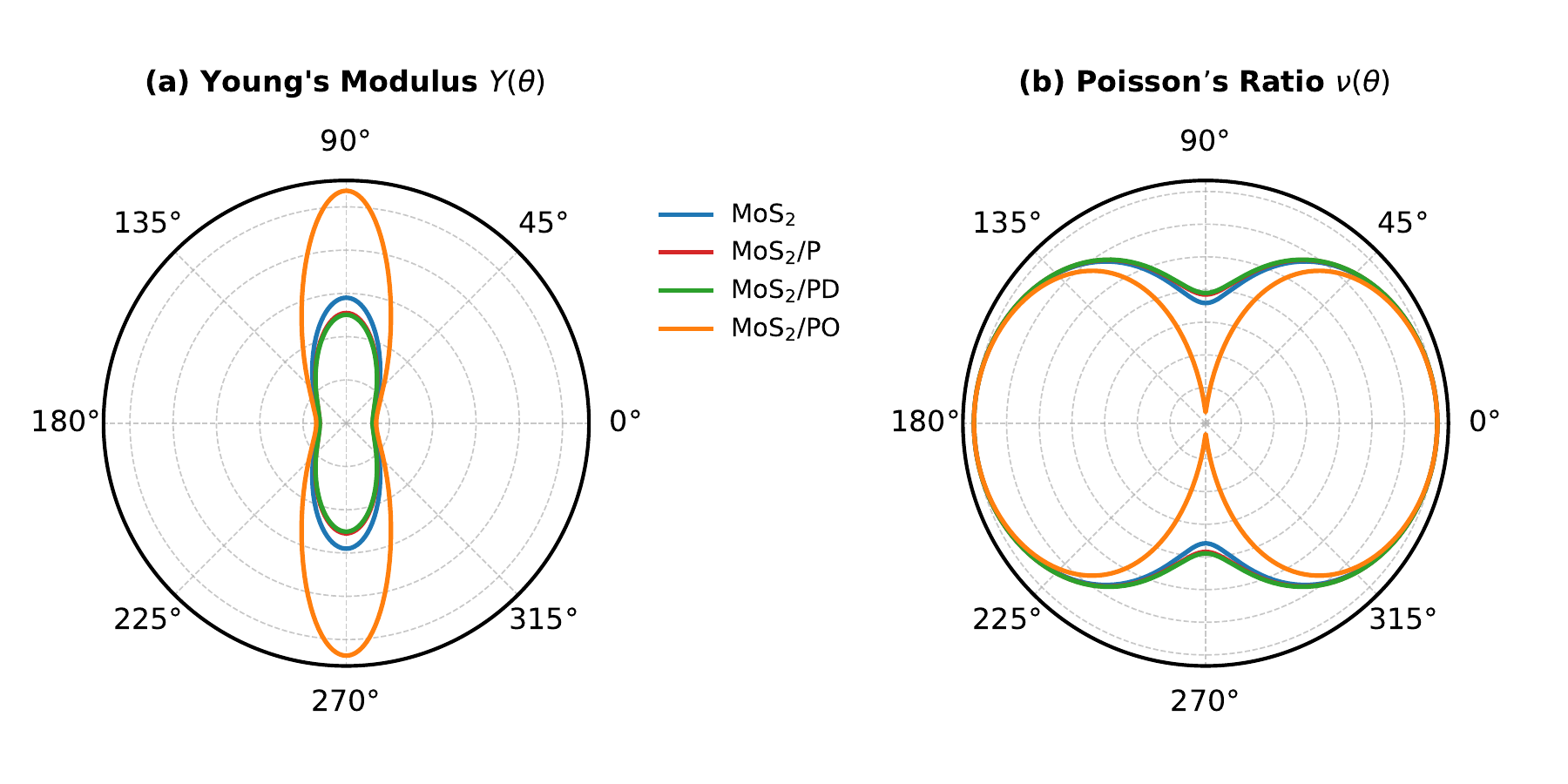}
  \caption{Elastic anisotropy of MoS$_2$/organic hybrid interfaces: (a) Polar plot of Young's modulus (eq.~\ref{eq:Y}) and (b) Poisson's ratio (eq.~\ref{eq:nu}) as functions of angular orientation. The angle $\theta$ is measured with respect to the $\vec{a}$ lattice axis ($\theta = 0^\circ/90^\circ$ corresponds to the armchair/zigzag direction). Contours in (a) are drawn in steps of 100~N/m starting from 100~N/m. Contours in (b) are drawn in steps of 0.1 starting from 0.}
  \label{polar_plots}
\end{figure}

\subsection{Mechanical Properties}

To assess the mechanical stability of the investigated hybrid systems,  elastic constants were computed.
The potential energy variation under uniaxial strain for these interfaces according to Eq. \ref{eq:es} is shown in the Suppl. Mat., Fig.~S2.  In the considered range of applied strain ($\pm 8$\%) all systems are in the elastic regime.
The results for the mechanical parameters are compiled in Table~\ref{tab1}. Note that the results for 
MoS$_2$ are consistent with values reported previously \cite{Dong.2017}. All systems satisfy Born’s stability criteria, specifically \(C_{11} > 0\) and \(|C_{11}| > |C_{12}|\), confirming their mechanical stability.

The in-plane stiffness, quantified by the 2D Young's modulus, shows significant variation across the systems, with  MoS$_2$/PO exhibiting the highest value (\(Y = 138.6 \, \rm{N/m}\)). MoS$_2$/PO also shows the lowest Poisson’s ratio (\(\nu = 0.13\)), indicating minimal transverse contraction under axial tension. In contrast, MoS$_2$/P and MoS$_2$/PD display slightly higher values, which may be related to differences in molecular geometry and intermolecular interactions rather than lattice compatibility, since no periodic or commensurate overlayer is formed. The bulk moduli of the three hybrid systems are approximately equal and smaller than for pristine MoS$_2$, i.e. physisorption reduces the resistance to isotropic in-plane deformation. The in-plane shear modulus is largest for MoS$_2$/PO. 

The different behavior of PO as compared to P and PD is also visible in the in-plane variations of Young’s modulus and Possion's ratio as seen in Fig.~\ref{polar_plots}(a,b). The anisotropic nature of Young’s modulus (panel (a)) is evident in all cases, with MoS$_2$/PO exhibiting the most pronounced deviations.  Concerning  Poisson’s ratio MoS$_2$/PO  again shows the largest anisotropy, exhibiting the lowest values at 90$^\circ$ and 270$^\circ$, corresponding to directions roughly perpendicular to the molecular axis. 

Rationalizing the difference between PO and P/PD in the magnitude of anisotropy one has to notice that not only is PO non-planar, but its distorted structure with respect to the gas phase introduces a considerable dipole moment of about 2~Debye (approximately along the long axis). Both give rise to a more pronounced directionality of the interaction with the MoS$_2$ surface, what is reflected in the response to lateral distortions  Notice that this is not reflected in the magnitude of the vdW binding energy (as compared to P and PD),  which refers to the perpendicular displacement.

\begin{figure}[h]
  \centering
  \includegraphics[width=0.8\textwidth]{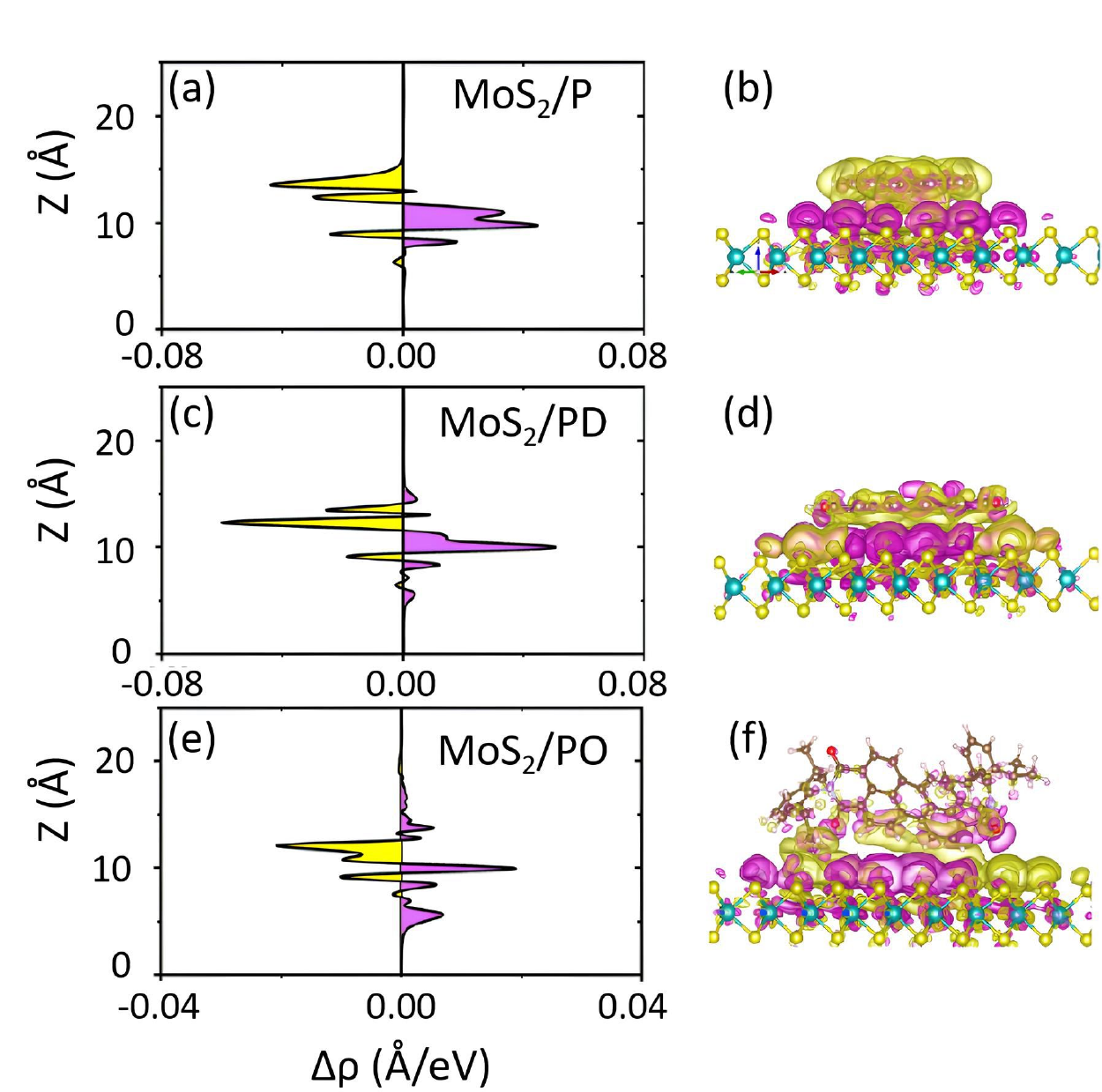}
  \caption{Planar-averaged charge density differences, $\Delta \rho(z)$, and isosurface of charge redistribution, $\Delta \rho({\bf r})$,  for the MoS$_2$/P (a,b), MoS$_2$/PD (c,d), and MoS$_2$/PO (e,f) hybrid interfaces. Panels (a), (c), and (e) depict $\Delta \rho(z)$ along the out-of-plane ($z$) direction. The gap between MoS$_2$ and the organic molecule is located around $z=11-12$~\AA. Panels (b), (d), and (f) show isosurface representations of the charge density difference with a contour threshold of ±0.002 e/\AA$^{3}$. 
  Interfacial polarization (dipolw formation) is indicated by charge depletion (hole-rich, purple) and charge accumulation (electron-rich, yellow).
}
  \label{fig:5}
\end{figure}

 \begin{figure}[h]
  \centering
  \includegraphics[width=0.9\textwidth]{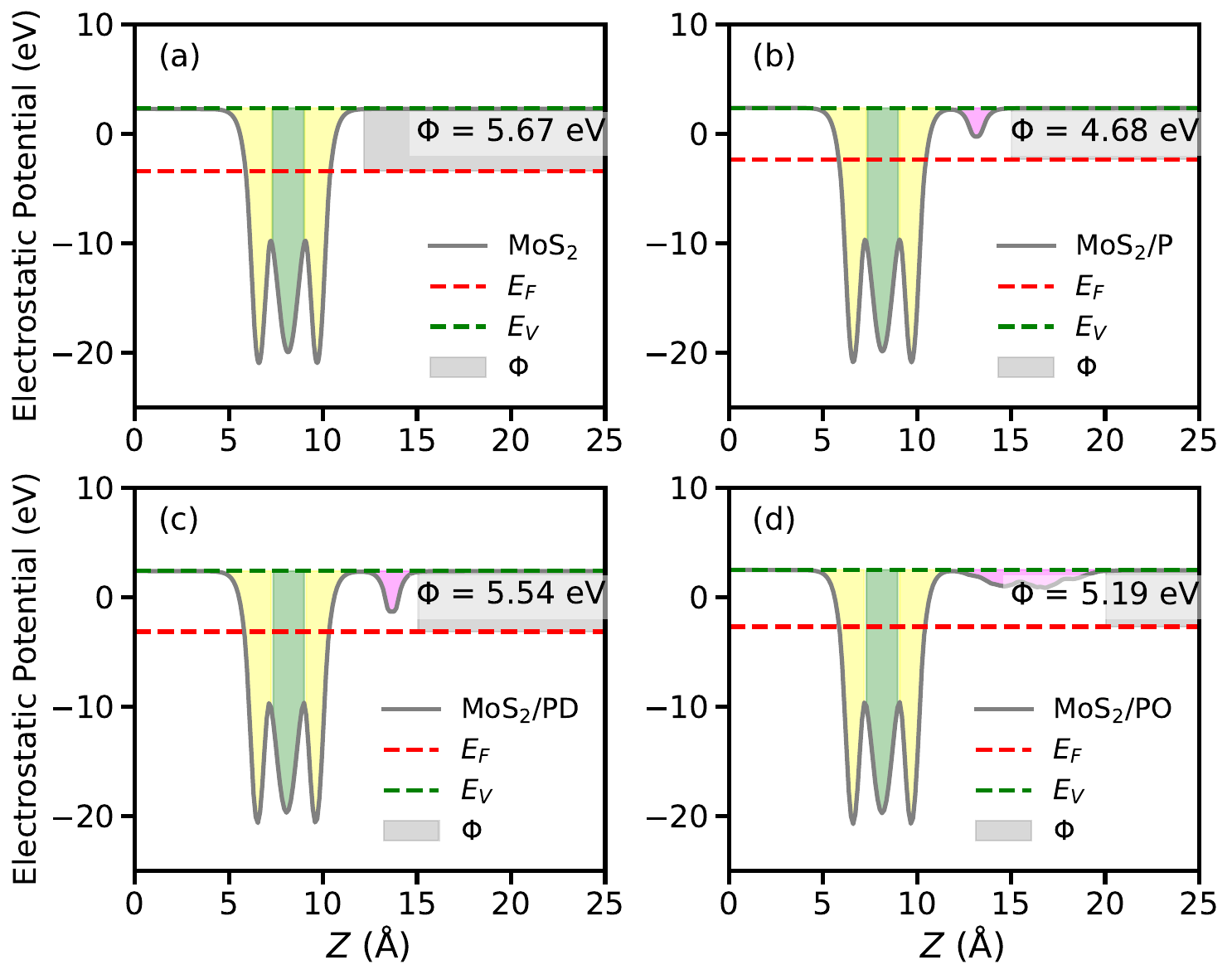}
  \caption{Planar-averaged electrostatic potential for  (a) MoS$_2$, (b) MoS$_2$/P, (c) MoS$_2$/PD, and (d) MoS$_2$/PO. The yellow and green regions represent the electrostatic potential contributions from the S and Mo atoms, respectively, while the purple region corresponds to the potential from the molecule.}
  \label{fig:4}
\end{figure}

\subsection{Interfacial Charge Transfer}

Charge redistribution at the MoS$_2$/organic interfaces was analyzed via charge density difference calculations, as shown in Fig.~\ref{fig:5}.
At the MoS$_2$/P interface (Fig.~\ref{fig:5}(a,b)), charge accumulation is predominantly localized on the planar P molecule, indicating its role as an electron acceptor, while hole accumulation occurs on the MoS$_2$ surface, confirming its function as an electron donor. The presence of the heteroatoms (O,N) leads to a charge transfer to the surface in the respective regions, yielding a more structured charge density difference distribution for PD and PO as seen in Fig.~\ref{fig:5}(c–f).  
Net charge transfer follow the trend \(\mathrm{MoS_2/PD} (0.013~e) > \mathrm{MoS_2/P} (0.011~e) > \mathrm{MoS_2/PO} (0.006~e)\). Notably it reflects the trend in binding energies.  The magnitude of charge transfer is very small, thus  confirming the physisorption character of the systems. In passing we note that this is in accord with results obtained for pentacene and PTCDA at MoS$_2$ \cite{Habib.2020}.

Charge transfer and interfacial polarization will also affect the work function. In Fig.~\ref{fig:4} the electrostatic potential profiles for pristine monolayer MoS$_2$ (Fig.~\ref{fig:4}(a)) and its interfaces with P (Fig.~\ref{fig:4}(b)), PD (Fig.~\ref{fig:4}(c)), and PO (Fig.~\ref{fig:4}d) are shown. For   MoS$_2$, the electrostatic potential curve displays pronounced minima near -20 eV, while the organic molecules exhibit smaller minima, generally around 0 eV to -2.5 eV. Clearly visible is the spread of charge density in case of the structurally distorted PO. Overall, the  notably lower electrostatic potential for MoS$_2$ relative to the organic layers, is confirming the observation of Fig. \ref{fig:5} that  MoS$_2$ functions as an electron donor. 

The pristine MoS$_2$ monolayer has a work function of 5.67 eV, in accord with previous results \cite{Habib.2020}.
The introduction of organic layers induces shifts in the work function to 4.70, 5.54, and 5.19 eV for P, PD, and PO, respectively. That is, the corresponding shifts are in the order P(-0.97~eV) $>$ PO -0.48~eV $>$ PD (-0.13~eV). In general, the observed shifts in work function upon organic layer addition are due to several factors, including interfacial dipole formation, band-bending effects, and substrate relaxation caused by adsorption.  
As these effects are not additive it is difficult to relate them separately to the work function change. For instance, the root mean squared deviation of the atomic positions with respect to pristine MoS$_2$ follows the order PO (0.28~\AA)$>$ P(0.158~\AA) $\sim$ PD(0.156~\AA). 
Comparing the present results with those of Ref. \cite{Habib.2020} we notice that work function changes are sizable and in between those of pentacene (-1.33~eV) and PTCDA (-0.036~eV).

\begin{figure*}[t]
  \centering
  \includegraphics[width=0.8\textwidth]{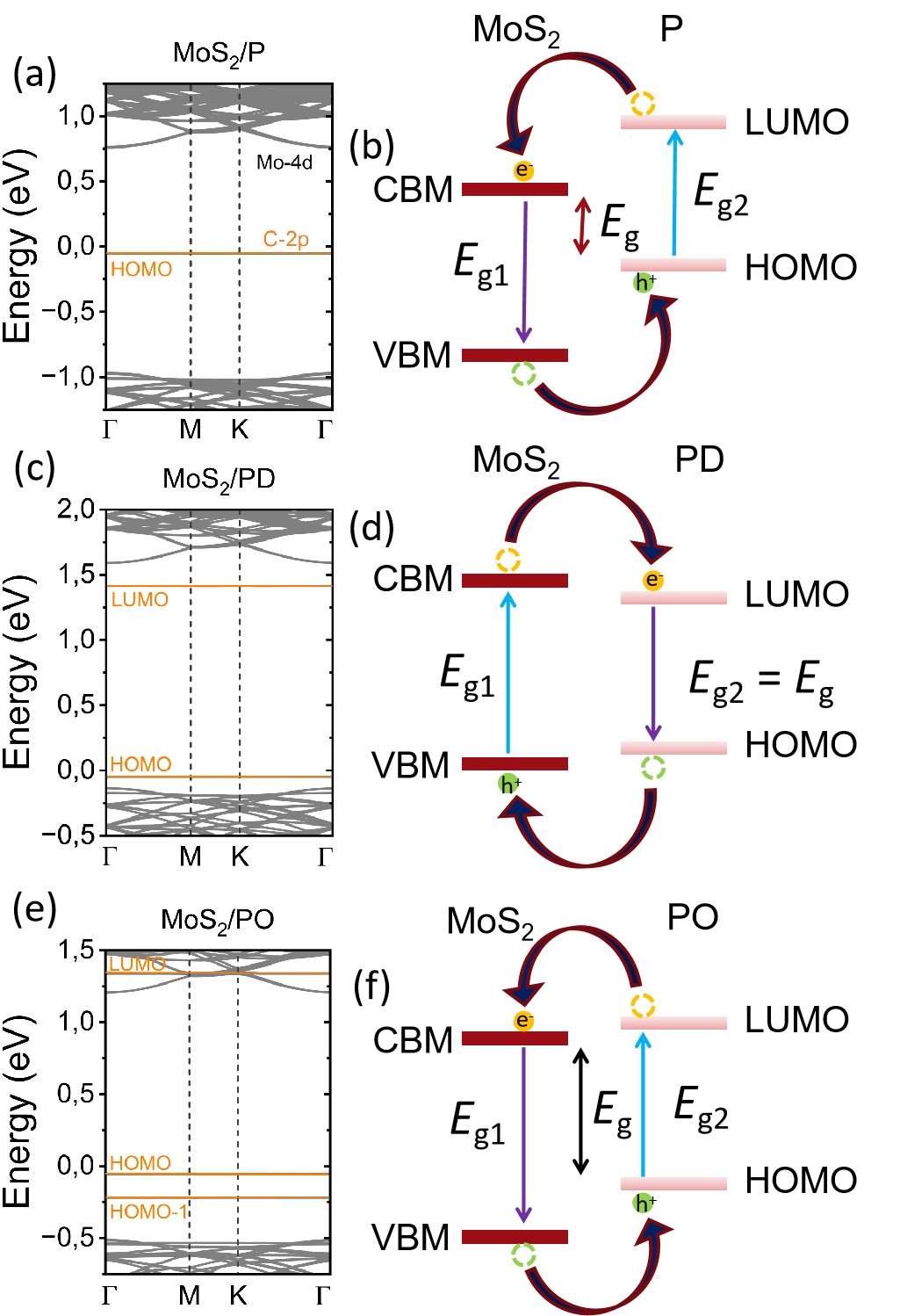}
  \caption{Band structures for the (a) MoS$_2$/P, (c) MoS$_2$/PD, and (e) MoS$_2$/PO hybrid interfaces (energies are given with respect to the Fermi energy $E_{\rm F}$). Panels (b), (d), and (f) show the corresponding schematic band alignment for each interface. The arrows indicate charge transfer processes that can occur upon photoexcitation of either P and PO (b,f) or MoS$_2$ (d).}
  \label{fig3}
\end{figure*}
\subsection{Band Structure}
\label{sec:band_structure}
\begin{figure}[h]
  \centering
  \includegraphics[width=0.9\textwidth]{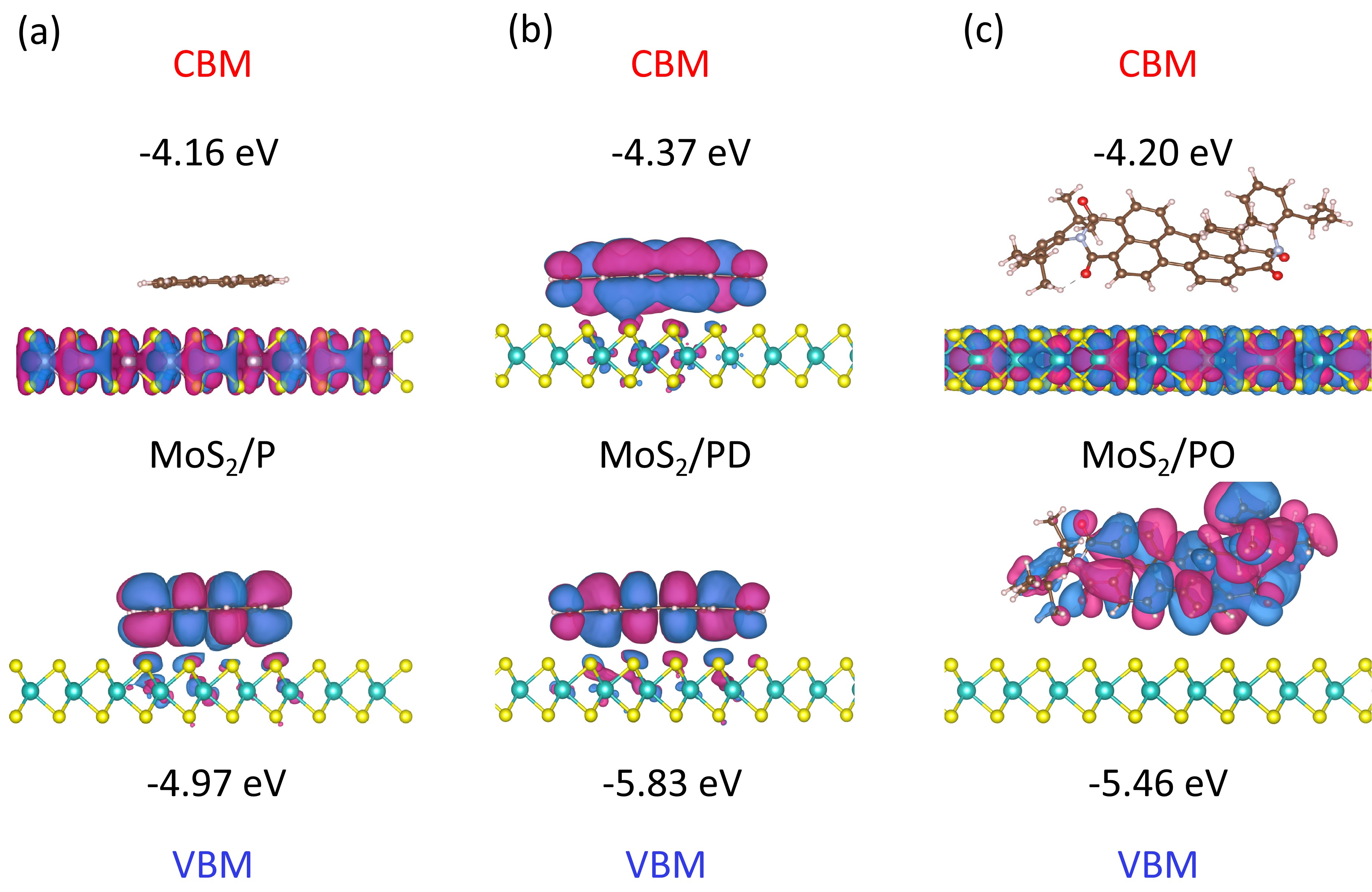}  
  \caption{Kohn–Sham wavefunctions at the VBM and CBM for (a) MoS$_2$/P, (b) MoS$_2$/PD, and (c) MoS$_2$/PO. The isosurfaces depict the spatial distribution of the wavefunctions, with blue (positive) regions indicating hole-like character and purple (negative) regions indicating electron-like character. Localization on both the MoS$_2$ layer and the organic molecule is evident. Numbers refer to orbital energies with respect to vacuum.}
  \label{fig:orbitals}
\end{figure}

The band structure and level scheme for the considered systems are shown in Fig.~\ref{fig3}. Using the PBE functional, the band gap of pristine MoS\(_2\) is found to be 1.65 eV, consistent with previous theoretical results \cite{krumland_2300089}. Introducing the organic molecules modifies the electronic structure, resulting in new states within the MoS\(_2\) band gap. This is reflected in the band structures shown in Fig.~\ref{fig3}(a), (c), and (e); the respective Kohn-Sham wavefunctions are given in Fig. \ref{fig:orbitals}.  For the partial density of states (PDOS), see Suppl. Mat., Figs.~S3–S5.

To investigate the extent to  which chemical modification of the organic part can be used to tune the hybrid interface, first we will focus on level energies with respect to vacuum for the constituents and the hybrid systems, cf. Tab. \ref{tab:levels}. Here, we notice that the conduction band minimum (CBM) and the valence band maximum (VBM) of MoS$_2$ are only marginally affected by the presence of the organic molecule.
For the organic molecules there are two effects, first the electronic interaction and second the changes geometry at the surface. Noticeable changes  of the LUMO (lowest unoccupied molecular orbital)/HOMO (highest occupied molecular orbital) energies due to modification of the geometry at the surface are observed for PO only. The additional effect due to the electronic interaction with the surface  is largest for P and moderate for PD and PO. Overall with respect to gas phase the HOMO-LUMO gap decreases by about 6\% for P and PD and increases by about 5\% for PO. In other words these energies are well preserved upon physisorption.

In each hybrid system, the interface exhibits, besides the overall band gap \(E_{\rm g}\), two distinct band gaps: \(E_{\rm g1}\), the difference between the CBM and VBM of MoS\(_2\), and \(E_{\rm g2}\), the difference between the LUMO and HOMO of the organic molecule. The different band gaps are summarized in Table~\ref{tab:bandgaps}.

\begin{table}[h]
\centering
\caption{Energy levels  for MoS$_2$/organic molecule hybrid and separate systems with respect to  vacuum level. Here X@MoS$_2$ refers to the isolated molecule having the geometry it will adopt at the surface.}
\begin{tabular}{l|cccc}
\hline\hline
System & VBM (eV) & CBM (eV) & HOMO (eV) & LUMO (eV)  \\
\hline
MoS$_2$          & –5.90 & –4.25 & &   \\
P     &   &   & –4.70  & –2.78  \\
PD     &   &   & -5.80   & -4.25   \\
PO     &   &   & –5.69  & –4.17 \\
P@MoS$_2$     &   &   & –4.70  & –2.78  \\
PD@MoS$_2$      &   &   & –5.80  & –4.29 \\
PO@MoS$_2$     &   &   & –5.48  & –4.05 \\
MoS$_2$/P      & –5.89 & –4.24 & –4.20 & –2.40 \\
MoS$_2$/PD    & –5.90 & –4.25 & –5.83 & –4.37  \\
MoS$_2$/PO     & –5.96 & –4.31 & –5.57 & –3.97  \\
\hline\hline
\end{tabular}
\label{tab:levels}
\end{table}

\begin{table}[h]
\centering
\caption{Calculated overall band gap \(E_{\rm g}\), MoS\(_2\) band gap \(E_{\rm g1}\), organic molecule band gap \(E_{\rm g2}\), and band alignment type for MoS\(_2\)/organic molecule hybrid systems.}
\begin{tabular}{l|lccc}
\hline\hline
System & \(E_{\rm g}\) (eV) & \(E_{\rm g1}\) (eV) & \(E_{\rm g2}\) (eV) & Band Alignment \\
\hline
MoS\(_2\)/P  & 0.81 (HOMO-CBM) & 1.65 & 1.80 & Type II \\
MoS\(_2\)/PD & 1.46 (HOMO–LUMO) & 1.65 & 1.46 & Type I \\
MoS\(_2\)/PO & 1.26 (HOMO–CBM) & 1.65 & 1.60 & Type II \\
\hline\hline
\end{tabular}
\label{tab:bandgaps}
\end{table}

In the MoS$_2$/P interface (Figs. \ref{fig3}(a) and \ref{fig:orbitals}(a)), the introduction of the P molecule creates localized states within the MoS$_2$ band gap. As shown in the band structure, the HOMO of P lies approximately in the middle of the MoS$_2$ band gap, while the  LUMO of P is positioned above the CBM of MoS$_2$. Thus we have a type II (staggered) band alignment.

For the MoS$_2$/PD interface, depicted in Figs.~\ref{fig3}(c) and and \ref{fig:orbitals}(b), both the HOMO and LUMO levels of PD lie within the band gap of MoS$_2$, with the HOMO positioned close to the VBM and the LUMO near the CBM of MoS$_2$. This energetic configuration corresponds to a type I (straddling) band alignment, where the organic molecule’s frontier orbitals are enclosed within the MoS$_2$ band gap. As a result, the effective band gap of the hybrid system (0.81~eV) is dominated by the HOMO–LUMO gap of PD.

Finally, the MoS$_2$/PO interface, shown in Figs.~\ref{fig3}(e) and \ref{fig:orbitals}(c), exhibits a type II (staggered) band alignment. Here, the HOMO and HOMO$-1$ levels of PO lie near the VBM of MoS$_2$, while the LUMO of PO is situated slightly above the CBM of MoS$_2$. The overall band gap of the MoS$_2$/PO system is 1.26~eV, defined by the energy difference between the PO HOMO and the MoS$_2$ CBM.

These findings are supported by the PDOS given in the Suppl. Mat., Figs.~S3–S5, which also illustrate the contributions of the different molecular orbitals. In the cases of P and PD the relevant HOMO/LUMO levels are of pure C-2p character. For PO the HOMO has additional contributions from O-2P and N-2p atomic orbitals. This distinct composition is reflected in the orbital plots in Fig. \ref{fig:orbitals}.
A summary of the type of band alignment is provided   in  Figs.~\ref{fig3}(b,d,f).

\section{Summary and Conclusions}

The objective in choosing perylene and its derivatives was to explore how the interfacial properties of the MoS$_2$/organic hybrid system change with the chemical modification of a given chromophore core. In terms of the optical absorption, PD and PO are similar having the lowest electronic transition around 2.3-2.4 eV. For P this transition is blue-shifted to about 2.8 eV. More relevant for the present discussion is the fact that  both P and PD are planar, the latter featuring heteroatoms (O,N) as well as a more extended $\pi$-electron system.  Compared to PD, PO has in addition bulky side groups attached to the N-sites along the long axis.  

All molecules are bound by vdW interaction to the MoS$_2$ surface. The binding energy of PD is larger than for P due the increased vdW contact with the surface. The bulky side groups of PO diminish the contact of its $\pi$-system with the surface resulting in the lowest binding energy. In addition the structures of PO is substantially distorted upon adsorption.

All systems show a pronounced mechanical anisotropy. However, PO stands out in terms of magnitude. A possible reason could be its distortion PO, leading not only to directionally sensitive steric effect but also causing a permanent dipole directed roughly along the long molecular axis.

Even though there is no appreciable hybridization of electronic orbitals, the interface is polarized, i.e. an interfacial dipole is formed. This comes along with a modification of the work function by up to about -1~eV for P as compared to bare MoS$_2$. The net charge transfer upon binding is on the order of 10$^{-2}$ to 10$^{-3}$ and reflects the order of binding energies. The pattern of charge density difference is shaped by the presence of the heteroatoms as well as by the structural distortion.

In terms of photophysical behavior band structure and interfacial level alignment is most important. For  MoS$_2$ the band gap is almost unaffected by the interaction with the organic molecules. The latter show some shift and also a net change of the HOMO-LUMO gap. For PO the largest change is due to the structural distortion, whereas for P one finds the largest effect of the electronic interaction. The band alignment is found to depend on the actual perylene, i.e. it is of type II for P and PO and of type I for PD. This has consequences for interfacial charge transfer pathways upon photoexcitation (cf. schemes in Fig. \ref{fig3}). Note that in case of PO the predicted band alignment is in accord with experimental PL measurements \cite{volzer23_3348a}. 

In summary, the present findings demonstrate the extent to which the interfacial properties of MoS$_2$/organic systems can be tuned through the chemical modification of a given chromophore. The chemical composition and geometry-enabled van der Waals contact play critical roles in the interfacial polarisation and band alignment of MoS$_2$/organic hybrid systems.

\ack
This work was funded by the Deutsche Forschungsgemeinschaft (DFG, German Research Foundation) - SFB 1477 "Light-Matter Interactions at Interfaces", project number 441234705”. 
\section*{References}
\bibliographystyle{iopart-num}
\bibliography{hybrid}

\end{document}